\documentclass[english,aps,manuscript]{revtex4}
\usepackage[T1]{fontenc}
\usepackage[latin1]{inputenc}
\usepackage{amssymb}

\makeatletter

\newcommand{\bee}{\begin{equation}}
\newcommand{\ee}{\end{equation}}
\newcommand{\beea}{\begin{eqnarray}}
\newcommand{\eea}{\end{eqnarray}}

\usepackage{babel}
\makeatother

\begin{document}

\preprint{COLO-HEP-528}

\maketitle
\begin{center}
\textbf{\Large On Anomaly Mediated SUSY Breaking}{\Large{}  }
\par\end{center}{\Large \par}

\begin{center}
\vspace{0.3cm}
 
\par\end{center}

\begin{center}
{\large S. P. de Alwis$^{\dagger}$ } 
\par\end{center}

\begin{center}
Physics Department, University of Colorado, \\
 Boulder, CO 80309 USA 
\par\end{center}

\begin{center}
\vspace{0.3cm}
 
\par\end{center}

\begin{center}
\textbf{Abstract} 
\par\end{center}

\begin{center}
\vspace{0.3cm}
 
\par\end{center}

A discrepancy between the Anomaly Mediated Supersymmetry Breaking
(AMSB) gaugino mass calculated from the work of Kaplunovsky and Louis
(hep-th/9402005) (KL) and other calculations in the literature is
explained, and it is argued that the KL expression is the correct
one relevant to the Wilsonian action. Furthermore it is argued that
the AMSB contribution to the squark and slepton masses should be replaced
by the contribution pointed out by Dine and Seiberg (DS) which has
nothing to do with Weyl anomalies. This is not in general equivalent
to the AMSB expression, and it is shown that there are models in which
the usual AMSB expression would vanish but the DS one is non-zero.
In fact the latter has aspects of both AMSB and gauge mediated SUSY
breaking. In particular like the latter, it gives positive squared
masses for sleptons.


\vfill{}

$^{\dagger}$ {\small e-mail: dealwiss@colorado.edu}{\small \par}

\eject

\section{Introduction}

Supersymmetry breaking is generally thought of as taking place in
a hidden sector and is communicated to the visible sector through
some messenger fields. The latter may be the moduli of string theory
which interact only with gravitational strength with the visible fields,
or some other messenger sector that couples to the gauge fields and
also to the supersymmetry breaking sector. The former mechanism may
be called moduli mediated supersymmetry breaking (MMSB) also known
as gravity mediated supersymmetry breaking (for a review see \citep{Nilles:1983ge}).
The latter is called gauge mediated supersymmetry breaking (GMSB -
see \citep{Giudice:1998bp} for a review). The advantage of the latter
over the former mechanism is that generically MMSB has flavor changing
(charge) neutral interactions and mass terms which need to be suppressed
by some fine tuning at the $10^{-3}$ level in order to agree with
experiment, while GMSB is naturally flavor neutral since the gauge
interactions are flavor blind.

An alternative to GMSB which shares its feature of being flavor blind
but like MMSB originates in the supergravity sector was proposed in
\citep{Randall:1998uk} \citep{Giudice:1998xp} \citep{Pomarol:1999ie}.
This mechanism has been called anomaly mediated supersymmetry breaking
(AMSB). This depends on the so-called Weyl (or conformal) anomaly
of supergravity (SUGRA) and appeared to depend on a particular formulation
of supergravity - namely the so-called Weyl (or conformal) compensator
formalism. This feature is rather puzzling and is clearly in need
of some explanation. In fact in \citep{Bagger:1999rd} there is an
argument based on the standard formulation of SUGRA, for the AMSB
gaugino masses, but not for the squark and slepton masses. In \citep{Dine:2007me}
on the other hand arguments are given for contributions to both gaugino
and scalar masses based on the need to preserve supersymmetry, independently
of any particular formulation of supergravity.

In this paper we will first argue that AMSB (for gaugino masses) is
in fact implicit in an old paper of Kaplunovsky and Louis \citep{Kaplunovsky:1994fg}%
\footnote{For earlier work see \citet{Dixon:1990pc} and \citet{Derendinger:1991hq}%
}. Two versions of the calculation were given there; one in the Weyl
compensator formalism and the other in the standard SUGRA formalism.
We show how the correct expression for the AMSB contribution to the
gaugino masses emerges from the compensator formalism. Then we rederive
this expression in the usual formulation (with the Weyl compensator
set to unity). In this case the contribution comes from Jacobians
in the measure coming from field redefinitions necessary to get to
the Kaehler-Einstein frame. The point of this is that the F-term of
the Weyl compensator is determined to have a value which is different
from that given in \citep{Randall:1998uk}\citep{Giudice:1998xp}
and \citep{Bagger:1999rd}%
\footnote{Bagger et al \citep{Bagger:1999rd} actually observe that there could
be additional contributions from high scale physics to their formula
for the gaugino mass. Also for related work see \citet{Gaillard:1999yb}. %
} and we discuss the reason for this difference and argue that the
correct contribution to the Wilsonian action is given (implicitly)
by \citep{Kaplunovsky:1994fg}. Next we discuss the contribution to
the gaugino masses pointed out by Dine and Seiberg \citep{Dine:2007me}(DS).
We argue that while it is certainly present, it is a new effect and
is not equivalent to the AMSB contribution. Finally we consider the
AMSB argument for soft masses. We point out that this actually violates
the Weyl invariance of this formulation of supergravity. Then we consider
the argument given in \citep{Dine:2007me} (DS). We generalize it
to show how a contribution to both Higgs sector and squark and slepton
sector masses can arise from this mechanism by following standard
supergravity calculations. We claim though that the DS mechanism is
a new one, i.e. is not equivalent to the AMSB argument, and in fact
(in the presence of a {}``mu'' term in the superpotential) gives
an additional term. We also point out that it is possible to find
models (see section IV) in which the usual AMSB expression vanishes
but the DS effect does not. Furthermore we argue that when the DS
effect is calculated by taking into account the fact that the wave
function renormalization at scales below the Higgs expectation value
depends on threshold effects, it is very similar to GMSB, and there
is no problem with the slepton masses. Of course unlike in GMSB the
gravitino mass is heavy, and sequestering \citep{Randall:1998uk}
is still necessary in order to ensure that the classical contribution
to the soft masses does not dominate the quantum effects. Finally
in an appendix we discuss in a simplified (non-supersymmetric) context
some issues relevant to Weyl transformations.

\section{Weyl Anomalies\label{sec:Weyl-Anomalies}}

The manifestly Weyl invariant formalism of $N=1$ supergravity (SUGRA)
is given by the following action (with $\kappa=M_{P}^{-1}=1,\, d^{8}z=d^{4}xd^{4}\theta,\, d^{6}z=d^{4}xd^{2}\theta$).\begin{eqnarray}
S & = & -3\int d^{8}z{\bf E}C\bar{C}\exp[-\frac{1}{3}K(\Phi,\bar{\Phi};Q,\bar{Q}e^{2V})]+\nonumber \\
 &  & \left(\int d^{6}z2{\cal E}[C^{3}W(\Phi,Q)+\frac{1}{4}f_{a}(\Phi){\cal W}^{a}{\cal W}^{a}]+h.c.\right)\label{eq:action}\\
 & = & -3\int d^{6}z2{\cal E}(-\frac{\bar{\nabla}^{2}}{4}+2R)C\bar{C}\exp[-\frac{1}{3}K(\Phi,\bar{\Phi};Q,\bar{Q}e^{2V})]+\nonumber \\
 &  & \left(\int d^{6}z2{\cal E}[C^{3}W(\Phi,Q)+\frac{1}{4}f_{a}(\Phi){\cal W}^{a}{\cal W}^{a}]+h.c.\right)\label{eq:action2}\end{eqnarray}
 In the above action $\Phi$, $Q$ are respectively a set of chiral
superfields representing the moduli and $ $the MSSM matter fields,
$V$is the gauge prepotential and ${\cal W}_{\alpha}=(-\frac{\bar{\nabla}^{2}}{4}+2R)e^{-2V}\nabla_{\alpha}e^{2V}$
is the associated gauge field strength. $R$ is the chiral curvature
superfield, ${\bf E}$ is the full superspace measure and ${\cal E}$
is the chiral superspace measure. The so-called torsion constraints
of SUGRA are invariant under Weyl transformations (with a chiral superfield
transformation parameter $\tau$) some of which are explicitly given
below.\begin{eqnarray}
{\bf E} & \rightarrow e^{2(\tau+\bar{\tau})}{\bf E}, & {\cal E}\rightarrow e^{6\tau}{\cal E},\nonumber \\
\nabla_{\alpha} & \rightarrow e^{(\tau-2\bar{\tau})}(\nabla_{\alpha}-\ldots), & V\rightarrow V,\nonumber \\
\Phi & \rightarrow\Phi,\, Q\rightarrow Q, & {\cal W}_{\alpha}\rightarrow e^{-3\tau}W_{\alpha}.\label{eq:weyl}\end{eqnarray}
 The Weyl compensator $C$ with the transformation rule\begin{equation}
C\rightarrow e^{-2\tau}C,\label{eq:Ctrans}\end{equation}
is then introduced in order to have a manifestly Weyl invariant action.
Note that since $C$ comes into the Kaehler potential in the form
$\ln C+\ln\bar{C}$ it is not a propagating field and the theory is
completely equivalent to the the usual formulation of supergravity.
However this will remain true for the quantum theory (meaning the
Wilsonian effective action rather than the 1PI effective action) only
to the extent that this Weyl invariance can be preserved. Any violation
of this invariance will result in the propagation of this degree of
freedom and hence produce a theory which is inequivalent to the original
supergravity. It should be stressed that provided supergravity is
not explicitly broken, the above formalism gives the most general
action, at the two derivative level, for a local supersymmetric theory
coupling pure supergravity to a locally gauge invariant theory of
chiral scalar superfields and gauge superfields.

The Weyl symmetry is anomalous at the quantum level because the path
integral measure is not invariant under these transformations. The
transformation of the measure can be obtained from the associated
chiral anomaly \citep{Gates:1983nr}\citep{Kaplunovsky:1994fg}\citep{Amati:1988ft}
and takes the form\begin{equation}
[d\Psi]\rightarrow[d\Psi]\exp\left\{ \frac{3c_{a}}{16\pi^{2}}\int d^{6}z2{\cal E}\tau{\cal W}^{a}W+h.c.\right\} .\label{eq:measure}\end{equation}
 Here the anomaly coefficient is given by\begin{equation}
c_{a}=T(G_{a})-\sum_{r}T_{a}(r)\label{eq:ca}\end{equation}
 and $T(G_{a}),T_{a}(r)$ are the trace of a squared generator in
the adjoint and the matter representation $r$ of the gauge group
$G_{a}$. For future use we also give here the 1-loop $\beta$-function
coefficient\begin{equation}
b_{a}=3T(G_{a})-\sum_{r}T_{a}(r).\label{eq:beta}\end{equation}
 This anomaly needs to be cancelled since the theory needs to retain
this local Weyl invariance and $C$ is a spurious degree of freedom.
This is done by the replacement \citep{Kaplunovsky:1994fg}\begin{equation}
f_{a}(\Phi)\rightarrow\tilde{f}(\Phi,C)\equiv f_{a}(\Phi)-\frac{3c_{a}}{8\pi^{2}}\ln C,\label{eq:ftilde}\end{equation}
 and it is easily seen from the transformation rules for $\Phi,C$,
that the anomaly is cancelled.$ $ This is essential (as stressed
in \citep{Kaplunovsky:1994fg}) in order to have equivalence between
the usual (manifestly supersymmetric) formulation of SUGRA where the
Weyl compensator is gauge fixed to $C=1$ and the Einstein-Kaehler
frame action (with Einstein gravity, canonical gravitino kinetic and
canonical Kaehler matter kinetic terms). The latter corresponds to
the gauge fixing\begin{equation}
\ln C+\ln\bar{C}=\frac{1}{3}K|_{Harm}\label{eq:Egauge}\end{equation}
 The instruction on the left hand side means that the chiral plus
anti-chiral pieces are to be taken (i.e. in components the lowest
component the $\nabla_{\alpha},-\frac{1}{4}\nabla^{2}$ and their
hermitian conjugates are to be retained %
\footnote{Note that this formula as well as similar formulae below leave the
phase of the lowest component undetermined. This means only that the
axionic partner of the gauge coupling has an ambiguity but it has
no effect on the gaugino masses.%
}). This is essentially the same as going to the Wess-Zumino gauge
for the hermitian superfield $K$.

Let us now expand the Kaehler potential in terms of the matter fields
by writing\begin{equation}
K(\Phi,\bar{\Phi};Q,\bar{Q}e^{2V})=K_{m}(\Phi,\bar{\Phi})+Z_{I\bar{J}}(\Phi,\bar{\Phi})\bar{Q}^{\bar{J}}e^{2V}Q^{I}+\ldots\label{eq:Kexpn}\end{equation}
 The first term in the action (\ref{eq:action}) then becomes\begin{equation}
\int d^{8}z{\bf E}C\bar{C}[-3e^{-\frac{1}{3}K_{m}(\Phi,\bar{\Phi})}+e^{-\frac{1}{3}K_{m}(\Phi,\bar{\Phi})}Z_{I\bar{J}}\bar{Q}^{\bar{J}}e^{2V}Q^{I}+\ldots].\label{eq:Acnexpn}\end{equation}
 To get canonical normalization for the matter fields we need to do
a field redefinition. For simplicity consider the case of one matter
field multiplet in a representation $r$. The kinetic term is contained
in \[
\int d^{8}z{\bf E}C\bar{C}e^{-K_{m}}Z_{r}(\Phi,\bar{\Phi})\bar{Q}e^{2V}Q,\, Z_{r}^{\dagger}=Z_{r}.\]
 Under the transformation $Q\rightarrow e^{\tau_{Z}}Q$ (where $\tau_{Z}$
is chiral) with all other fields fixed, the path integral measure
acquires a factor\begin{equation}
\exp\left\{ -\frac{T_{a}(r)}{16\pi^{2}}(\int d^{6}z2{\cal E}\tau_{Z}{\cal W}^{a}{\cal W}^{a}+h.c.)\right\} \label{eq:canmeasure}\end{equation}
 This implies that under this transformation the gauge coupling function
in the quantum action becomes\begin{equation}
H_{a}(\Phi,C,\tau_{Z})\equiv\tilde{f}_{a}(\Phi,C)-\frac{T_{a}(r)}{4\pi^{2}}\tau_{Z}=f_{a}(\Phi)-\frac{3c_{a}}{8\pi^{2}}\ln C-\frac{T_{a}(r)}{4\pi^{2}}\tau_{Z}\label{eq:ftilde2}\end{equation}
 and the matter kinetic term becomes\[
\int d^{8}z{\bf EC\bar{C}}e^{-\frac{1}{3}K_{m}}Ze^{\tau_{Z}+\bar{\tau}_{Z}}\bar{Q}e^{2V}Q.\]
 To get canonical normalization for the matter kinetic term we need
to put\begin{equation}
\tau_{Z}+\bar{\tau}_{Z}=\ln(C\bar{C}e^{-\frac{1}{3}K_{m}}Z_{r})|_{Harm}\label{eq:tauZconstr}\end{equation}
 where the instruction on the right hand side means that the equality
holds only for its harmonic part. Defining\[
h_{a}=H_{a}|,\, h_{aR}=\Re H_{a}|,\]
 the gauge coupling and the gaugino mass are given by (see for example
\citep{Wess:1992cp} equation (G.2)) $\frac{1}{g_{a}^{2}}=h_{aR}$
and\begin{equation}
m_{a}=h_{aR}^{-1}\Re(F^{i}\partial_{i}h_{a}+F^{C}\partial_{C}h_{a}+F^{\tau_{Z}}\partial_{\tau_{Z}}h_{a}).\label{eq:gauginomass}\end{equation}
 Using (\ref{eq:ftilde2}) and (\ref{eq:tauZconstr}) we then have
\begin{eqnarray}
\frac{1}{g_{a}^{2}} & = & h_{aR}=(\Re f(\Phi)-\frac{b_{a}}{16\pi^{2}}\ln(C\bar{C})-\frac{T_{a}(r)}{8\pi^{2}}\ln(e^{-\frac{1}{3}K_{m}}Z_{r})|\label{eq:coupling}\\
 & = & \Re f(\Phi)|-\frac{c_{a}}{16\pi^{2}}K_{m}|-\frac{T_{a}(r)}{8\pi^{2}}\ln Z_{r}|\end{eqnarray}
 The last expression is valid in the Einstein frame and we used (the
lowest component of) (\ref{eq:Egauge}) to obtain it. This is of course
exactly the expression given in \citep{Kaplunovsky:1994fg} (see equation
C.15) evaluated at the cutoff scale and ignoring the term proportional
to $\ln\Re f_{a}$ which comes from rescaling to get the canonical
kinetic term for the gauge potential. It should stressed here that
that in \citep{Kaplunovsky:1994fg} these expressions were also evaluated
directly by explicit computations which showed that they are independent
of whether or not a manifestly supersymmetric regulator was used,
and confirmed the argument using the Weyl anomaly. Note also that
the various scalar fields are to be evaluated at the minimum of the
potential and in particular we have assumed that MSSM fields $Q$
are set to zero at this point (so that $K|_{0}=K_{m}|_{0}$ for instance).
The formula can be easily corrected if some of these fields are Higgses
which have non vanishing vacuum expectation values. Similarly the
gaugino masses are given by\begin{eqnarray}
\frac{m_{a}}{g_{a}^{2}} & = & \Re[F^{i}\partial_{i}f_{a}(\Phi)|-\frac{b_{a}}{8\pi^{2}}\frac{F^{C}}{C}-\frac{T_{a}(r)}{4\pi^{2}}F^{i}\partial_{i}(\ln(e^{-\frac{1}{3}K_{m}}Z_{r})]\label{eq:gauginomass1}\\
 & = & \Re[F^{i}\partial_{i}f_{a}(\Phi)|-\frac{c_{a}}{8\pi^{2}}F^{i}\partial_{i}K_{m}-\frac{T_{a}(r)}{4\pi^{2}}F^{i}\partial_{i}(\ln Z_{r})]\label{eq:gauginomass2}\end{eqnarray}
 The sum over $i$ it should be recalled goes over all the moduli
(which are of course gauge neutral) and in the general case of more
than one matter representation a sum over $r$ is implied. Also to
go from the first line to the second in the above expression we used
the F-component of (\ref{eq:Egauge}). The F-component of the moduli
fields are as usual given by the formula \begin{equation}
F^{i}=-e^{K/2}K^{i\bar{j}}D_{\bar{j}}\bar{W}.\label{eq:Fi}\end{equation}
 At this point it behooves us to explain the differences between (\ref{eq:gauginomass2})
and what has appeared before in the literature. In \citep{Randall:1998uk}
and \citep{Giudice:1998xp} it is asserted that $F^{C}/C=m_{3/2}$,
whereas here (following \citep{Kaplunovsky:1994fg}) it is fixed by
the Einstein-Kaehler gauge condition (\ref{eq:Egauge}). In \citep{Bagger:1999rd}
the formula that is given for the gaugino mass is (after adding the
classical piece to equation (4) of that paper and changing the normalizations
to agree with ours)\begin{equation}
\frac{m_{a}}{g_{a}^{2}}=\Re[F^{i}\partial_{i}f_{a}(\Phi)|-\frac{1}{8\pi^{2}}(b_{a}m_{3/2}+c_{a}F^{i}\partial_{i}K_{m}+2T_{R}F^{i}\partial_{i}\ln Z_{r})],\label{eq:Bagger}\end{equation}
 (though as the authors observed the calculation is sensitive to high
scale effects). This formula could be obtained from our formula (\ref{eq:gauginomass1})
if instead of using (the F-term of) equation (\ref{eq:Egauge}) we
used the formula\begin{equation}
\frac{F^{C}}{C}=m_{3/2}+\frac{1}{3}F^{i}K_{i}.\label{eq:FCnew}\end{equation}
 In order to understand the meaning of one choice over the other it
is instructive to first consider the equation of motion for the $C$
field. Take the second form of the action (\ref{eq:action2}) and
vary it with respect to $C$ to get \begin{equation}
(-\frac{\bar{\nabla}^{2}}{4}+2R)\bar{C}\exp[-\frac{1}{3}K(\Phi,\bar{\Phi};Q,\bar{Q}e^{2V})]+C^{2}W=0\label{eq:Ceom}\end{equation}
 Taking the lowest component of this equation and taking the value
of $C|$ from (\ref{eq:Egauge})) we get (ignoring fermionic terms)\begin{equation}
\frac{\bar{F}^{\bar{C}}}{\bar{C}}+2R|=e^{K/2}W|+\frac{1}{3}\bar{F}^{\bar{i}}K_{\bar{i}}\label{eq:FCbyC}\end{equation}
 So in the Einstein-Kaehler gauge, i.e. using the F-component of (\ref{eq:Egauge}),
this equation just determines the (lowest component) of the chiral
curvature, $2R|=e^{K/2}W|=m_{3/2}$. The equation (\ref{eq:FCnew})
would be compatible with the equation of motion for $C$ only in a
gauge in which $R|=0$. $ $

The fact that the correct value of the AMSB contribution to the gaugino
mass is given by (\ref{eq:gauginomass2}) can also be seen in a different
way - one that does not depend on the Weyl invariance argument of
Kaplunovsky and Louis \citep{Kaplunovsky:1994fg} and would be equivalent
to the alternate argument given there %
\footnote{It should be pointed out that the absence of the $m_{3/2}$ term of
(\ref{eq:Bagger}) as in (\ref{eq:gauginomass2}), has been shown
in an explicit string theory calculation in \citet{Antoniadis:2005xa}.%
}. In other words we will just use the standard supergravity formulation
which corresponds to taking the gauge $C=1$ in (\ref{eq:action}).
In this case to get to the Einstein-Kaehler gauge we need to make
a Weyl transformation (\ref{eq:weyl}) with \begin{equation}
2\tau+\bar{2\tau}=\frac{K_{m}}{3}|_{Harm}.\label{eq:taunew}\end{equation}
 From (\ref{eq:measure}) we see that this is tantamount to making
the replacement\[
f_{a}(\Phi)\rightarrow f_{a}(\Phi,\tau)=f_{a}(\Phi)-\frac{3c_{a}}{4\pi^{2}}\tau.\]
 The matter kinetic terms are now $\int d^{8}z{\bf E}Z_{r}(\Phi,\bar{\Phi})\bar{Q}e^{2V}Q$.
Next we need to do redefine the matter fields $Q$ to get canonical
normalization for them. This corresponds to the transformation $Q\rightarrow e^{\tau_{Z}}Q$
with\begin{equation}
\tau_{Z}+\bar{\tau}_{Z}=\ln(Z_{r})|_{Harm}.\label{eq:tauznew}\end{equation}
 Again there is a contribution from the measure - namely (\ref{eq:canmeasure}),
so that the effective gauge coupling function is finally\begin{equation}
H_{a}(\Phi,\tau,\tau_{Z})=f_{a}(\Phi)-\frac{3c_{a}}{4\pi^{2}}\tau-\frac{T_{a}(r)}{4\pi^{2}}\tau_{Z}.\label{eq:ftautauz}\end{equation}
 Using (\ref{eq:taunew})(\ref{eq:tauznew}) and taking the F-component
we again get (\ref{eq:gauginomass2}).

It should be stressed that this contribution to the gauge coupling
function has nothing to do with renormalization group running and
the beta-function. The formulae (\ref{eq:coupling})(\ref{eq:gauginomass2})
are statements about the theory at the Wilsonian cutoff (say $\Lambda$)
where $f_{a}$ is defined as the gauge coupling function in the original
SUGRA frame. If we change the cutoff (say from $\Lambda$ to $\mu$
then to one-loop order we have\begin{equation}
H_{a}(\Phi,\tau,\tau_{Z},\mu)=H_{a}(\Phi,\tau,\tau_{Z},\Lambda)-\frac{b_{a}}{8\pi^{2}}\ln\frac{\Lambda}{\mu}\label{eq:RG}\end{equation}
 It should be noted that while this last term contributes to the evolution
of the gauge coupling, the ratio of the gaugino mass to the squared
coupling is independent of the running since the RG running term is
a constant and only contributes to the lowest component of the superfield
gauge coupling (however see \citep{Dine:2007me} and the discussion
below).

Of course as with all our previous considerations (\ref{eq:gauginomass1})(\ref{eq:gauginomass2})(\ref{eq:ftautauz})
are valid only for the Wilsonian coupling function which is not renormalized
beyond one loop. This function however is not the physical coupling
since the kinetic terms for the gauge fields is not canonically normalized.
In order to get the physical coupling %
\footnote{We are essentially following an argument due to Arkani-Hamed and Murayama
\citep{ArkaniHamed:1997mj} in the global case and we give this here
for completeness, even though it is not germane to our main considerations.%
} we need to make a further transformation by a chiral superfield $\tau_{v}$,
\begin{equation}
V=e^{(\tau_{V}+\bar{\tau}_{V})/2}V_{c},\label{eq:Vcan}\end{equation}
 such that the gauge field term in the action\begin{equation}
\frac{1}{4}\int d^{6}z{\cal E}\tilde{H}_{a}{\cal W}^{a}(e^{\tau_{V}+\bar{\tau}_{V})/2}V_{c}){\cal W}^{a}(e^{\tau_{V}+\bar{\tau}_{V})/2}V_{c})\label{eq:Vckinetic}\end{equation}
 is canonically normalized. Here we have redefined $H_{a}$ to include
a term coming from the measure so that $\tau_{V}$ is to be determined
from the equation\begin{equation}
\Re\tilde{H}_{a}\equiv\Re H_{a}-\frac{T_{a}(G_{a})}{8\pi^{2}}2\Re\tau_{V}=e^{-2\Re\tau_{V}}\equiv\frac{1}{g_{phys}^{2}},\label{eq:gphys}\end{equation}
 so that the gauge field kinetic terms have canonical normalization.
Combining the equations (\ref{eq:taunew}, \ref{eq:tauznew}, \ref{eq:ftautauz},
\ref{eq:RG}, \ref{eq:gphys}) gives us the NSVZ equation \citet{Novikov:1985rd}
for the physical coupling in a locally supersymmetric theory.

Now let us comment on the calculation of \citep{Bagger:1999rd} which
is done in the $C=1$ gauge. This is based on the 1PI effective action
of \citep{LopesCardoso:1993sq} where the non-local term \begin{eqnarray}
\Delta L & = & -\frac{g^{2}}{(16\pi)^{2}}\int d^{2}\theta2{\cal E}{\cal W}{\cal W}\frac{4}{\square}(-\frac{\bar{\nabla}^{2}}{4}+2R)\nonumber \\
 &  & \{b_{a}4\bar{R}+\frac{T_{a}(r)}{3}\nabla^{2}K+T_{a}(r)\nabla^{2}\ln Z_{r}\}+h.c.\label{eq:Cardoso}\end{eqnarray}
 is added. Here $\square$ is the flat-space Laplacian. This non-local
action is designed to reproduce the super-Weyl anomalies that we have
discussed and it is globally supersymmetric but is not locally supersymmetric
(it is actually not generally covariant). Such a non-local action
could have local ambiguities which need to be fixed by some criterion.
The value of the gaugino mass coming from (\ref{eq:Cardoso}) is what
is given in \citep{Bagger:1999rd} and quoted in equation (\ref{eq:Bagger}).
In fact there is a simpler way of deriving this same result - with
a similar problem. Thus instead of just adding the $-\frac{3c_{a}}{8\pi^{2}}\ln C$
term as in \citep{Kaplunovsky:1994fg} to cancel the anomaly, one
again works in the $C=1$ gauge and adds a term \begin{equation}
-\frac{3c_{a}}{8\pi^{2}}\ln\phi\label{eq:lnphi}\end{equation}
 where $\phi={\cal E}^{1/3}$, to \textit{reproduce} the Weyl anomaly
\footnote{In fact the argument of \citep{Giudice:1998xp} is similar to this.%
}. Then in (\ref{eq:gauginomass1}) the $\ln C$ term would be replaced
by a $\ln\phi$ term. Then noting that (since $-\nabla^{2}{\cal E}/4|=6e\bar{R}|$
(see for example \citep{Wess:1992cp} equations (20.21,22)) we have\begin{equation}
\frac{F^{\phi}}{\phi}=2\bar{R}=m_{3/2}+\frac{1}{3}F^{i}K_{i},\label{eq:Fphi/phi}\end{equation}
 where the last equation is equation (\ref{eq:FCbyC}) in the $C=1$
Weyl gauge. In other words the effect of replacing $C$ by $\phi$
is to use (\ref{eq:FCnew}) in (\ref{eq:gauginomass1}) giving us
(\ref{eq:Bagger}) as we argued earlier. However $\phi$ unlike $C$
is not really a chiral scalar. Although it is chiral, $\phi^{3}={\cal E}$
is a chiral density and so the term we added, like the non-local action
of \citep{LopesCardoso:1993sq} but unlike the term $\ln C$, is globally
but not locally supersymmetric. Also this term gives an unusual term
proportional to $\ln e$ in the expression for the coupling. This
clearly shows that we have introduced a diffeomorphism anomaly though
of course in flat space it is zero. Similarly the non-local addition
(\ref{eq:Cardoso}) gives a non-local contribution to the gauge coupling.

We conclude that the correct anomaly mediated contribution to the
gaugino mass in the Wilsonian action is given by (\ref{eq:gauginomass2}).
In fact as we showed in the discussion leading to (\ref{eq:ftautauz})
the calculation just depends on using the appropriate expressions
for the relevant Jacobians in going to the Einstein-Kaehler frame
with canonical normalization for the matter fields, and is completely
unambiguous.

However an additional contribution to the gaugino mass arises from
an effect first noticed by Dine and Seiberg \citep{Dine:2007me}(DS).
This is usually ignored since the vacuum expectation values of the
MSSM fields are set to zero. However some of these fields (Higgses)
have non-zero expectation values in the physical vacuum and these
authors propose that in effect the RG scale $\mu^{2}$ should be replaced
by $\chi_{+}\chi_{-}$ where $ $$\chi_{\pm}$ are a pair of Higgs
fields (for instance they could be the MSSM charge neutral Higgs superfields
$h_{u,d}^{0}$ which have equal and opposite hypercharge). In fact
this is what should be done in a background field calculation of the
one-loop effective action. In this case the gauge coupling function
$H$ at the MSSM scale would have an additional term\begin{equation}
H_{a}\sim\frac{b_{a}}{16\pi^{2}}\ln\frac{\chi_{+}\chi_{-}}{\Lambda^{2}}\label{eq:HDS}\end{equation}
 To preserve supersymmetry $\chi_{\pm}$ must be the complete superfield.
Of course as pointed out in \citep{Dine:2007me} this formula is only
valid in the Higgs phase of the theory. This then gives an additional
contribution to the gaugino mass\begin{equation}
\frac{m_{a}}{g_{a}^{2}}\sim\frac{b_{a}}{16\pi^{2}}\Re(\frac{F^{+}}{\chi^{+}}+\frac{F^{-}}{\chi^{-}})\label{eq:maDS}\end{equation}
 We emphasize that this expression gives an unambiguous contribution
to the gaugino mass since we are in the Higgs phase. Of course in
the symmetric phase this expression would be of the form 0/0 and ambiguous,
but in this phase eqn (\ref{eq:HDS}) would no longer be valid and
one would need an explicit infra-red cutoff. In the MSSM for example
this effect is present only in the physical Higgs vacuum.

To see what this contribution is in a concrete example note that after
the various field redefinitions discussed earlier the MSSM fields
have canonical normalization and in particular we may take (setting
the Planck scale $M_{P}=1$) \begin{equation}
K\sim\chi_{+}\bar{\chi}_{+}+\chi_{-}\bar{\chi}_{-}+Q\bar{Q}\ldots,\, W\sim W_{0}+m\chi_{+}\chi_{-}+h\chi_{+}Q^{2}.\label{eq:toymssm}\end{equation}
The ellipses in $K$ represent the hidden sector fields and $W_{0}$
is the superpotential in the hidden sector with the hidden sector
fields having a supersymmetry breaking minimum at some low scale generating
a non-zero gravitino mass $m_{3/2}=e^{K_{0}/2}W|_{0}$ . This is of
course just a toy version of the MSSM with $Q$ being the {}``top''
quark/squark (with {}``hypercharge'' $-\frac{1}{2}$) whose loops
can induce gauge symmetry breaking in the usual fashion (see for example
\citep{Weinberg:2000cr}, \citep{Drees:2004jm}\citep{Baer:2006rs}).
The actual situation in the MSSM is in fact a straightforward generalization
of this. Thus after hidden sector supersymmetry breaking this model
will be in the Higgs phase so that (\ref{eq:HDS})(\ref{eq:maDS})
make sense and are unambiguous. As in the MSSM vacuum then $\chi_{\pm}=v_{\pm}\ne0,$
which may without loss of generality be chosen real (as in the MSSM)
and $Q=0$. Defining $\frac{v_{+}}{v_{-}}=\tan\beta$ we get \begin{equation}
\bar{F}^{\pm}|_{0}=e^{K|_{0}/2}(mv_{\mp}+v_{\pm}W|_{0}).\label{eq:Fplusminus}\end{equation}
 Defining $\tilde{m}=e^{K_{0}/2}m$ we get from (\ref{eq:maDS}) the
contribution\begin{equation}
\frac{m_{a}}{g_{a}^{2}}\sim\frac{b_{a}}{8\pi^{2}}(m_{_{3/2}}+\tilde{m}{\rm cosec}2\beta).\label{eq:maDS1}\end{equation}
 If one ignored the {}``mu'' term contribution (i.e. the second
term in the paranthesis), it would seem that we have restored the
$O(m_{3/2})$ present in (\ref{eq:Bagger}). However the origin of
these terms is very different. Furthermore as we will see in the next
section the interpretation of $\chi^{\pm}$ as Higgs superfields will
result in a further modification which will result in a formula analogous
to the one in GMSB%
\footnote{In \citep{Dine:2007me} actually the vevs $v_{\pm}$ are  taken to
zero in which case, in the absence of a {}``mu'' term the results
will be completely equivalent to those of AMSB, including of course
the unfortunate negative slepton mass squared result. For more details
see section \ref{sec:DS-susy-breaking}.%
}. Thus we conclude that the above effect is a new one which adds to
the AMSB effect, which as argued earlier is actually given by (\ref{eq:gauginomass2})
rather than (\ref{eq:Bagger}). In fact as is evident from the above
calculation it depeds on the form of the visible sector superpotential.
If there is no {}``mu'' term as in the example considered by Dine
and Seiberg \citep{Dine:2007me} then the contribution is the same
as that in the old AMSB calculations such as that of \citep{Bagger:1999rd}.
But if there is a {}``mu'' term (as there must be in any realistic
theory of low energy supersymmetry) then there is another term coming
from the DS calculation that is not present in the old AMSB calculations.

\section{Soft masses in AMSB\label{sec:Soft-masses-in}}

In addition to a contribution to the gaugino mass, AMSB effects are
supposed to contribute to the soft masses of MSSM scalar fields as
well as to their couplings. Let us first review the usual argument.
This may be motivated from the following observation for the gauge
coupling superfield chiral scalar function $H_{a}$. Using the Weyl
compensator formalism the Wilsonian coupling at some scale $\Lambda$
can be written (by combining (\ref{eq:ftilde2}) and (\ref{sec:Soft-masses-in})
as \begin{equation}
H_{a}(\Phi,C,\tau_{Z})=f_{a}(\Phi)-\frac{b_{a}}{8\pi^{2}}\ln C-\frac{T_{a}(r)}{4\pi^{2}}\ln(e^{-\frac{1}{3}K_{m}}Z_{r})\label{eq:HC}\end{equation}
 where it is implied that only the lowest (whose phase is undetermined)
and $\theta$ and $\theta^{2}$ components of the last term are taken.
The gauge coupling function at some scale $\mu$ is then given by
adding the term $-\frac{b_{a}}{8\pi^{2}}\ln\frac{\Lambda}{\mu}$ (as
in (\ref{eq:RG})) giving the the coupling function at scale $\mu$
as \begin{equation}
H_{a}(\Phi,C,\tau_{Z})_{\mu}=f_{a}(\Phi)-\frac{b_{a}}{8\pi^{2}}\ln(C\Lambda/\mu)-\frac{T_{a}(r)}{4\pi^{2}}\ln(e^{-\frac{1}{3}K_{m}}Z_{r}).\label{eq:HCmu}\end{equation}
 This might lead to the supposition that one should replace $\Lambda/\mu$
by $C\Lambda/\mu)$ in the superfield functions that occur in the
Wilsonian action evaluated at the scale $\mu$. In particular the
wave function renormalization function $Z(\Phi,\bar{\Phi},\ln\frac{\Lambda}{\mu})$
at scale $\mu$ might be replaced by $Z(\Phi,\bar{\Phi},\ln\frac{\Lambda|C|}{\mu})$.
If this is indeed justified then there would be an anomaly mediated
contributions to the soft masses \citep{Randall:1998uk}\citep{Pomarol:1999ie},\begin{equation}
m^{2}=-\ln Z|_{\theta^{2}\bar{\theta}^{2}}=-\frac{1}{4}|F^{C}|^{2}\frac{d^{2}\ln Z}{d\ln\Lambda^{2}}.\label{eq:softmass}\end{equation}
 This would be the dominant contribution if the usual classical contribution
(from the F-term of $\Phi$ is suppressed by sequestering (see \citep{Randall:1998uk}).
However the origin of the $\ln C$ term and the $\ln\Lambda/\mu$
terms in (\ref{eq:HCmu}) is completely different. The first exists
even without any running i.e already at the high scale where the classical
coupling is defined. It comes from the field redefinition/Weyl transformation
Jacobians/anomalies. The second is a consequence of running. More
importantly if one used the function $Z(\Phi,\bar{\Phi},\ln\frac{\Lambda|C|}{\mu})$
in the Wilsonian action then it is no longer invariant under the Weyl
transformations and hence it would not be possible to remove $C$
from the theory. In fact it is precisely the Weyl variation of the
$\ln(C\Lambda/\mu)$ term in (\ref{eq:HCmu}) that guarantees the
$ $Weyl invariance of the quantum theory by canceling the Weyl anomaly.
Finally the derivation of the gauge coupling function in the Einstein-Kaehler
frame given in the discussion from (\ref{eq:taunew}) to (\ref{eq:ftautauz}),
shows that the extra terms in the gauge coupling function are just
a consequence of the field redefinitions. The apparent symmetry between
the $\ln C$ term and the RG term $\ln\Lambda/\mu$ term has no physical
significance. From the Wilsonian point of view the two scales $\Lambda$
and $\mu$ are both physical scales and should be measured in the
same conformal gauge. Thus their ratio should be independent of the
conformal gauge that is chosen. Indeed if one works in the $C=1$
gauge one can still derive the contribution to the gauge coupling
function as we did above, but there is no field corresponding to $C$
that can be used since the only other possibility, namely $\phi$,
is not really a chiral scalar but a density, and as we pointed out
earlier its use would violate local supersymmetry/general covariance.

The problem with the usual AMSB argument is that it is based on conformal
invariance rather than Weyl invariance. Unlike the conformal invariance
the Weyl invariance exists whether or not there are mass terms. This
is because it involves transforming the metric whereas in the usual
discussion Weyl invariance is spontaneously broken by restricting
the argument to flat space. If this is done one loses sight of the
(super) general covariance of the supergravity action. In other words
the invariance in question involves transforming the background that
is held fixed in the usual discussion. If the Weyl invariance of the
action is violated (as would be the case if $C$ dependence is introduced
into the wave function renormalization then local supersymmetry will
not be preserved.

An alternative mechanism for generating soft masses was given in \citep{Dine:2007me}
(DS). The mechanism is quite general but let us first discuss it within
the context of the example given in that paper. The supergravity potentials
are given by the following. \begin{eqnarray}
K & = & -3\ln[1-\frac{1}{3}K_{v}(\chi,\bar{\chi})-\frac{1}{3}K_{h}(z,\bar{z}))],\label{eq:K}\\
K_{h} & = & z\bar{z}-\frac{z^{2}\bar{z}^{2}}{\mu^{2}},\label{eq:K_h}\\
K_{v} & = & Z(\chi\bar{\chi})\chi\bar{\chi},\label{eq:K_V}\\
Z & = & 1+\epsilon a_{1}\ln(|\chi|^{2}/\Lambda^{2})+\epsilon^{2}a_{2}\ln^{2}(|\chi|^{2}/\Lambda^{2}),\label{eq:Z}\\
W & = & W_{0}-M^{2}z+W_{v}(\chi),\label{eq:W}\end{eqnarray}
 with $M_{P}=1,\, M\ll\mu\ll1$ $\epsilon=g^{2}/16\pi^{2}$ and $a_{1,2}$
are model dependent numbers. The constant $W_{0}$ is tuned such that
$V_{0}=0$ and at the minimum we have (ignoring the matter sector)\begin{equation}
F^{z}\simeq M^{2},\, m_{3/2}=M^{2}/\sqrt{3},\, z=z_{0}\equiv\mu^{2}/2\sqrt{3}\ll1.\label{eq:min}\end{equation}
 The visible sector is assumed to be such that at the minimum of the
combined potential $F^{\chi}\ll M^{2}$ and $\chi_{0}\ll z_{0}$.
We also have near the minimum\begin{eqnarray}
K & \simeq & K_{v}+K_{h}+\frac{1}{3}K_{v}K_{h}+\ldots,\, K_{z\bar{z}}\simeq1+O(\mu^{2}),\, K_{\chi\bar{\chi}}\simeq1+O(\epsilon)\nonumber \\
K_{v\chi} & \simeq & \bar{\chi},\, K_{h\bar{z}}\simeq z,\, K_{\chi\bar{z}}\simeq\frac{1}{3}\bar{\chi}z=-K^{\chi\bar{z}}.\label{eq:Kmetric}\end{eqnarray}
 With $\partial_{i}V_{0}=V_{0}=0,$ the (squared) soft mass is essentially
the Fermi-Bose splitting of the squared masses and is given by (see
for example \citep{Wess:1992cp} p187-188) \begin{eqnarray}
\Delta m_{\chi\bar{\chi}}^{2} & = & M_{\chi\bar{\chi}}^{2}-m_{\chi\bar{\chi}}^{2}=e^{K_{0}}[-R_{\chi\bar{\chi}k\bar{l}}K^{k\bar{m}}K^{\bar{l}n}D_{n}WD_{\bar{m}}W+K_{\chi\bar{\chi}}|W|^{2}]\nonumber \\
 & = & e^{K_{0}}[-(R_{\chi\bar{\chi}z\bar{z}}(K^{z\bar{z}})^{2}|D_{z}W|^{2}+R_{\chi\bar{\chi}\chi\bar{\chi}}(K^{\chi\bar{\chi}})^{2}|D_{\chi}W|^{2})\nonumber \\
 &  & +K_{\chi\bar{\chi}}|W|^{2}+O(\mu^{2}m_{3/2}^{2})].\label{eq:DeltaM2}\end{eqnarray}
 In standard calculations of soft mass terms (see for example \citep{Kaplunovsky:1993rd})
only the first term in the second line above is kept since SUSY breaking
happens in the hidden sector and $|D_{\chi}W|_{0}=0$. However here
the Kaehler metric is singular at $\chi=0$$ $, so there are extra
terms if $|D_{\chi}W|$ goes to zero no faster than linearly.$ $
We find\[
R_{\chi\bar{\chi}z\bar{z}}\simeq\frac{1}{3}K_{hz\bar{z}}K_{v\chi\bar{\chi}},\, R_{\chi\bar{\chi}\chi\bar{\chi}}\simeq K_{\chi\bar{\chi}}(2a_{2}-a_{1}^{2})\frac{\epsilon^{2}}{\chi\bar{\chi}}.\]
 As expected (since $K$ is of the sequestered form) the usual contribution
vanishes. So we get (since $K_{\chi\bar{\chi}}=1+O(\epsilon)$ and
$e^{K_{0}}\simeq1$) for the normalized soft mass squared,\begin{eqnarray}
m_{s}^{2} & \simeq & -R_{\chi\bar{\chi}\chi\bar{\chi}}|F^{\chi}|^{2}=-R_{\chi\bar{\chi}\chi\bar{\chi}}|K^{\chi\bar{\chi}}|^{2}|D_{\chi}W|^{2}\nonumber \\
 & \simeq & (a_{1}^{2}-2a_{2})\frac{\epsilon^{2}}{\chi\bar{\chi}}|\partial_{\chi}W_{v}+\bar{\chi}W|^{2}=\epsilon^{2}(a_{1}^{2}-2a_{2})|m_{3/2}+O(\frac{\partial_{\chi}W_{v}}{\bar{\chi}})|^{2},\label{eq:m_sds}\end{eqnarray}
 where in the last two steps we used (\ref{eq:min}) and (\ref{eq:Kmetric}).
Note that all classical contributions to the soft masses cancel because
of the sequestered form of the Kaehler potential . If there are no
{}``mu'' terms (i.e. terms of the form $m\chi^{2}$) in $W_{v}$
then we have the result of DS.

Let us compare this to the usual AMSB formula (\ref{eq:softmass}).
If we assume that its F-term is given by (\ref{eq:FCnew})\begin{equation}
\frac{F_{C}}{C}=m_{3/2}+\frac{1}{3}F^{z}\partial_{z}K=m_{3/2}(1+O(\mu^{2})).\label{eq:FC}\end{equation}
 Also \begin{equation}
\gamma=-\partial\ln Z/\partial\ln\Lambda=2\epsilon a_{1}+4\epsilon^{2}(2a_{2}-a_{1}^{2})\ln\frac{|\chi|}{\Lambda},\label{eq:gamma}\end{equation}
 where in the last step we used (\ref{eq:Z}). This then gives\begin{equation}
m_{s}^{2}=\epsilon^{2}(a_{1}^{2}-2a_{2})m_{3/2}^{2}\label{eq:m_samsb}\end{equation}
 in agreement with the DS calculation if there are no {}``mu'' terms.
However the appearance of a {}``mu'' term contribution in the DS
calculation means that it is not completely equivalent to AMSB. For
instance if the {}``mu'' term is fine-tuned to cancel exactly the
$\bar{\chi}W$ term (at the minimum) so that $D_{\chi}W$ vanished
quadratically with $\chi$, the DS contribution would be absent. Finally
the AMSB argument for scalar masses would involve breaking the Weyl
invariance while the DS calculation does not.

The above is valid for the Higgs fields of the low-energy theory but
it is not clear from the above how the squarks and sleptons (which
should have zero expectation values) should get DS type contribution
to their mass. To see how this happens let us extend the DS toy model
by adding a superfield $Q$ (standing for a toy version of a quark
or lepton superfield) which will have zero expectation value and no
mass term but having a Yukawa interaction with the {}``Higgs'' field
$\chi$. Thus we replace the matter Kaehler potential by

\begin{equation}
K_{v}=Z(\chi\bar{\chi})(\chi\bar{\chi}+Q\bar{Q})\label{eq:kv1}\end{equation}
 where $Z$ is again given by (\ref{eq:Z}) and write the superpotential
$W_{v}$ in(\ref{eq:W}) as\begin{equation}
W_{v}=\frac{m}{2}\chi^{2}+h\chi Q^{2}+\ldots\label{eq:Wvnew}\end{equation}
 with the ellipses containing terms which are higher order in the
fields. Now we assume that the latter are such that the potential
has a minimum (see also the discussion in the previous section) with
\begin{equation}
\chi_{0}=\bar{\chi}_{0}=v,\, Q_{0}=0.\label{eq:vev}\end{equation}
 From the above we have $F^{\bar{\chi}}=(\tilde{m}+m_{3/2})v,\, R_{Q\bar{Q}\chi\bar{\chi}}=\epsilon^{2}(2a_{2}-a_{1}^{2})/v^{2}$.
Then following the same steps as in (\ref{eq:DeltaM2})(\ref{eq:m_sds})
we get \begin{equation}
\Delta m_{Q\bar{Q}}^{2}=-R_{Q\bar{Q}\chi\bar{\chi}}|F^{\chi}|^{2}=\epsilon^{2}(a_{1}^{2}-2a_{2})(\tilde{m}+m_{3/2})^{2}.\label{eq:Deltamqqbar}\end{equation}
 Thus we do indeed have a contribution to the soft masses but again
as was case with the DS contribution to the gaugino masses, it has
nothing to do with Weyl anomalies. Furthermore unlike what is usually
claimed as a contribution to the scalar mass from AMSB, the DS contribution
fits naturally into the standard calculation of soft mass terms in
supergravity.

\section{Models with $F^{C}=0$ and non-zero DS effect\label{sec:Models-with-F^{C}=00003D00003D0}}

As noted in \citep{Randall:1998uk}, for the dominant contribution
to the soft masses to be from AMSB the classical contribution from
SUSY breaking in the hidden sector needs to be sequestered - as in
equation (\ref{eq:K}) above. The same is obviously true for the alternative
to AMSB, namely the DS version discussed above. A sequestered version
that can naturally arise in type IIB string theory is one of the GKP-KKLT
\citep{Giddings:2001yu} \citep{Kachru:2003aw} type with the visible
sector being on a set of D3 branes. In such a model with just one
Kaehler modulus $T$ acquiring a non-zero F-term at the minimum (it
is fine tuned by choosing fluxes and non-perturbative terms so that
the cosmological constant is zero and SUSY breaking is only from this
modulus) the soft masses will indeed be zero, and both the so-called
$A$ and $B$ terms are also zero.

We first consider here the simplest version of this - namely the so-called
no-scale model (which in type IIB is derived by GKP \citep{Giddings:2001yu}).
This illustrates the point, although of course the scale of SUSY breaking
and the modulus $T$ are not fixed. As is well known the soft masses
and the $A$ and the $B$ terms, are all zero in such models (see
for example \citep{Brignole:1997dp} and references therein) even
though supersymmetry is broken with a non-zero gravitino mass and
a zero cosmological constant.

The point that we want to illustrate here is that when calculating
the soft masses, the appropriate input from supergravity has to be
taken for $F^{C}$. Thus consider the following toy model for the
superpotential and Kaehler potential.\begin{eqnarray}
W & = & W_{mod}+mH^{2}\label{eq:toyW}\\
K & = & -3\ln(T+\bar{T}-\frac{1}{3}H\bar{H})\simeq K_{mod}+ZH\bar{H}+O(H^{2}\bar{H}^{2})\label{eq:toyK}\end{eqnarray}
 with $K_{mod}=-3\ln(T+\bar{T})$ and $Z=1/(T+\bar{T})$ and $\partial_{T}W_{mod}=0$.
The standard argument in supergravity consists of evaluating the usual
expression for the potential for the chiral scalars and then extracting
the scalar mass terms i.e. coefficients of the $H\bar{H},\, HH$ terms,
and one finds that they are zero.

What would the corresponding effective global calculation yield. In
computing the potential one can of course ignore the chiral curvature
($R$) terms and effectively work with the flat space Lagrangian \begin{eqnarray}
L & = & -3\int d^{4}\theta C\bar{C}e^{-K/3}+(\int d^{2}\theta C^{3}W+h.c.)\label{eq:flatL}\\
 & = & -3\int d^{4}\theta C\bar{C}e^{-K_{mod}/3}+(\int d^{2}\theta C^{3}W_{mod}+h.c.)\label{eq:flatsugra}\\
 & + & \int d^{4}\theta\hat{H}\bar{\hat{H}}+(\int d^{2}\theta Cm\hat{H}^{2}+h.c.)\label{eq:global}\end{eqnarray}
 In the last two lines we have used the above toy model and rescaled
(as is usual in AMSB type calculations) $H\rightarrow\hat{H}\equiv CH$.
Now the usual discussion of AMSB proceeds from the last line. If this
were whole story (as far the visible sector were concerned) there
would be for instance a problem with the so-called $B\mu$ term i.e.
the coefficient of the $H^{2}$ (where $H$ refers to the scalar component)
in the potential. For this would be then given by (see for example
the review \citep{Luty:2005sn}) $F^{C}m$. However the value of $F^{C}$
needs to be fixed from the line (\ref{eq:flatsugra}) of this equation.
In fact of course the first line (\ref{eq:flatL}) leads (upon elimination
of $C$) to the usual SUGRA potential and therefore to the result
that all soft terms are zero.

Obviously one should get the same result from the second form (\ref{eq:flatsugra})
plus (\ref{eq:global}) of the Lagrangian. In this version the line
(\ref{eq:flatsugra}) is used to get $F^{C}$ (up to small corrections
$O(H^{2})$ and this SUGRA input must be used to compute effects in
the `MSSM' sector of line (\ref{eq:global}). So from (\ref{eq:flatsugra})
we get as usual from the (lowest components of) the equations of motion
for $C$ and the chiral super fields,\begin{eqnarray}
\bar{F}^{\bar{C}} & = & C^{2}e^{K/3}(W-\frac{1}{3}K_{\bar{j}}K^{\bar{j}i}D_{i}W)\label{eq:FCgeneral}\\
 & = & C^{2}e^{K_{mod}/3}(W_{mod}-\frac{1}{3}\frac{3}{T+\bar{T}}\frac{(T+\bar{T})^{2}}{3}\frac{3W_{mod}}{(T+\bar{T})})+O(H^{2})=O(H^{2}),\label{eq:FCnoscale}\end{eqnarray}
 since $D_{T}W_{mod}=\partial W_{mod}-3W_{mod}/(T+\bar{T})=-3W_{mod}/(T+\bar{T})$
in this no-scale case. Thus the $B\mu$ term is actually zero (to
$O(H^{2})$) as are all other soft terms.

A similar situation exists for more realistic models where the $T$
modulus is stabilized. The MSSM sector will have (schematically) quark/lepton
superfields denoted by $Q$ and Higgs fields denoted by $H$. For
the Kaehler potential we take\begin{eqnarray}
K & = & -3\ln(T+\bar{T}-\frac{1}{3}(H\bar{H}+Q\bar{Q}))-\ln(S+\bar{S})+k(z,\bar{z})\label{eq:Kfull}\\
 & = & K_{mod}+Z(H\bar{H}+Q\bar{Q)}+\ldots\\
K_{mod} & = & -3\ln(T+\bar{T})-\ln(S+\bar{S})+k(z,\bar{z}),\, Z=\frac{1}{T+\bar{T}}.\label{eq:KmodZ}\end{eqnarray}
 For the moduli superpotential we take a GKP-KKLT \citep{Giddings:2001yu,Kachru:2003aw}
form\begin{equation}
W_{mod}=W_{flux}(S,z)+\sum_{n}A_{n}(S,z)e^{-a_{n}T},\label{eq:W0new}\end{equation}
 while for the 'MSSM' superpotential we take \[
W_{MSSM}=mH^{2}+yHQQ\]
 In the above $S$ is the dilaton-axion superfield and $z=\{z^{r}\}$
denotes the set of complex structure moduli and $T$ is the Kaehler
modulus of some Calabi-Yau orientifold (with $h_{11}=1$) compactification
of type IIB string theory. Such a model can be realized as a generalization
of those considered by GKP-KKLT \citep{Giddings:2001yu}\citep{Kachru:2003aw}.
Also the MSSM sector is located on a stack of D3 branes. The moduli
potential is then\begin{equation}
V_{mod}=\frac{e^{k(z,\bar{z})}}{(S+\bar{S})(T+\bar{T})^{2}}\{\frac{1}{3}|\partial_{T}W_{mod}|^{2}-2\Re\partial_{T}W_{mod}\bar{W}_{mod}\}+|F^{S}|^{2}K_{S\bar{S}}+F^{z}F^{\bar{z}}k_{z\bar{z}}.\label{eq:Vmoduli}\end{equation}
 Now one looks for a local minimum of this potential with zero cosmological
constant (CC) and SUSY breaking only in the $T$ direction, i.e. \begin{equation}
V_{mod}|_{0}=0,\, F|_{0}^{S}=F^{z}|_{0}=0,\, F|_{0}\ne0.\label{eq:susybreakmin}\end{equation}
 There is certainly no obstruction to finding such a minimum and with
a sufficient number of complex structure moduli and non-perturbative
terms it is reasonable to expect that such a SUSY breaking minimum
exists. The fine tuning condition for the CC now takes the form (at
the above local minimum of the potential)\begin{equation}
|D_{T}W_{mod}|_{0}^{2}\frac{(T+\bar{T})_{0}^{2}}{3}=3|W_{mod}|_{0}^{2},\label{eq:finetune}\end{equation}
 or taking the same phase as in the no-scale model we have\begin{equation}
D_{T}W_{mod}|_{0}(T+\bar{T})_{0}^{}=-3W_{mod}|_{0}.\label{eq:finetune2}\end{equation}
 It should be stressed that unlike in the case of the no-scale model
(where these relations are automatic) in the present case they are
fine tuned relations that are valid at the SUSY breaking local minimum
(\ref{eq:susybreakmin}). In effect the relation implies that we should
fine tune such that $\partial_{T}W|_{0}=0$ which is certainly possible
if there are at least two non-perturbative terms in (\ref{eq:W0new}).
Using (\ref{eq:FCgeneral}) and (\ref{eq:finetune2}) we get \begin{equation}
F^{C}|_{0}=C^{2}e^{K_{mod}/3}(W_{mod}+D_{T}W|_{mod}\frac{T+\bar{T}}{3})|_{0}=0,\label{eq:FCzero}\end{equation}
 ignoring $O(H^{2})$ terms. Defining $\hat{H}=CH,\,\hat{Q}=CQ$ as
in (\ref{eq:global}) we see that the effective `MSSM' theory is given
by (noting that $e^{-K_{mod}/3}Z=(S+\bar{S})^{1/3}k^{-1/3}(z,\bar{z})$)\begin{equation}
L_{MSSM}=\int d^{4}\theta(S+\bar{S})^{1/3}k^{-1/3}(z,\bar{z})(\hat{H}\bar{\hat{H}}+\hat{Q}\bar{\hat{Q}})+\{\int d^{2}\theta(mC\hat{H}^{2}+y\hat{H}\hat{Q^{2}})+h.c.\}.\label{eq:MSSMnew}\end{equation}
 Since this is independent of the SUSY breaking modulus $T$, and
as we saw above $F^{C}$ is also zero at the minimum of the moduli
potential, all soft SUSY breaking terms are zero.

In addition to the vanishing of the classically generated soft terms,
in this model the usual AMSB expression is also zero. The latter is
obtained by inserting a wave function renormalization factor $Z(\mu C/\Lambda$)
into the first term of (\ref{eq:MSSMnew}) and this gives a contribution
to the soft terms proportional to $F^{C}$. But since the latter is
zero at the minimum of the moduli potential in this model, there is
no such contribution. Nevertheless the DS mechanism gives a non-zero
contribution. This arises from a wave function renormalization factor
$Z(H\bar{H})$ and the soft terms are proportional to $|\frac{D_{H}W}{H}|_{0}=|m+W_{mod}\frac{\bar{H}}{H}|$
which is generally non-zero. This clearly illustrates the fact that
the DS mechanism is not equivalent to the usual AMSB argument.

\section{DS SUSY breaking and GMSB\label{sec:DS-susy-breaking}}

In the section \ref{sec:Soft-masses-in} we rederived the DS formula
for the soft masses and showed that it is different from the AMSB
one if there is a mu-term. Here we will revisit the calculation and
argue that it needs to be seriously modified when the field $\chi$
is identified with the Higgs field. The reason is that the scale of
the soft masses is around the scale of the Higgs vacuum expectation
value. This will lead us to conclude that the problem of negative
squared slepton masses that plagues AMSB is absent in the DS mechanism.

Let us first briefly review the calculation of soft masses in GMSB
using the method of \citet{Giudice:1997ni}. Defining $\alpha=g^{2}/4\pi$
where $g$ is the coupling of some gauge group, the anomalous dimension
of some chiral scalar field $Q$ is given (to one loop order) by\begin{equation}
\gamma\equiv\frac{d\ln Z}{d\ln\mu}=\frac{c}{\pi}\alpha.\label{eq:gamma1}\end{equation}
 Here $Z$ is the wave function renormalization of $Q$ at the scale
$\mu$ (so that the Kaehler potential for it is $Z(\mu)Qe^{V}\bar{Q}$)
and $c=c_{2}(r)$ is the quadratic Casimir for the representation
$r$. Suppose that between the ultraviolet scale $\Lambda$ (which
could be the Planck scale or the scale associated with the hidden
sector where SUSY is broken) there is an intermediate scale (messenger
mass in GMSB) characterized by a chiral scalar superfield $X$ (which
can develop a non-zero vacuum expectation value (vev) and an F-term).
Thus $X=\chi$ of section III or $X=\sqrt{\chi^{+}\chi^{-}}$ of section
II and in the MSSM should be taken to be the invariant $X=\sqrt{H^{u}H^{d}}$.
The beta function well above and well below the scale set by the vev
of $X$ are given to one loop by $\beta'=-b'g^{3}/16\pi^{2},\,\beta=-bg^{3}/16\pi^{2}$.
Integrating these last two equations we have

\begin{eqnarray}
\alpha_{X}^{-1} & = & \alpha_{\Lambda}^{-1}+\frac{b'}{4\pi}\ln\frac{X\bar{X}}{\Lambda^{2}}\label{eq:alphaX}\\
\alpha_{\mu}^{-1} & = & \alpha_{X}^{-1}+\frac{b}{4\pi}\ln\frac{\mu^{2}}{X\bar{X}}\label{eq:alphamu}\end{eqnarray}
 Note that the coupling at the low scale $\mu$ depends on the threshold
scale $X$. In a supersymmetric theory the scale $X$ should be replaced
by the complete superfield and $\alpha^{-1}$ is the real part of
the chiral superfield $f$. Integrating (\ref{eq:gamma1}) then gives\begin{equation}
\ln Z(\mu)=\ln Z(\Lambda)+\frac{2c}{b'}\ln\frac{\alpha_{\Lambda}}{\alpha_{X}}+\frac{2c}{b}\ln\frac{\alpha_{X}}{\alpha_{\mu}}\label{eq:logZ}\end{equation}
 The radiatively generated soft mass is given by\begin{equation}
m_{Q}^{2}(\mu)=-\frac{\partial^{2}\ln Z(\mu)}{\partial\ln X\partial\ln\bar{X}}\frac{|F_{X}|^{2}}{|X|^{2}}\label{eq:m2Qmu}\end{equation}
 where it is understood that the right hand side is evaluated at the
vev of $X$. Using (\ref{eq:logZ}) to evaluate the derivatives and
taking the limit $\mu\rightarrow X$ (where now $X$ is the vacuum
expectation value of the field) we have\begin{equation}
m_{Q}^{2}(X)=2c\left(\frac{\alpha_{X}}{4\pi}\right)^{2}(b-b')\frac{|F_{X}|^{2}}{|X|^{2}}=2c\left(\frac{\alpha_{X}}{4\pi}\right)^{2}(b-b')m_{3/2}^{2},\label{eq:m2X}\end{equation}
 where in the last step we have identified $X$ with the field $\chi$
of section \ref{sec:Soft-masses-in} and ignored the 'mu' term. The
beta function above the threshold has more matter states contributing
than the one below, so $b-b'$ is always positive and (\ref{eq:m2X})
implies that (since $c>0$) the squared masses are always positive
as in GMSB. Let us contrast this with the calculation that was done
in section \ref{sec:Soft-masses-in}, and see why it needed to be
modified. There what was done was in effect to first take the limit
$\mu\rightarrow X$ in (\ref{eq:logZ}) in which case the last term
in the right hand side of that equation disappears. Then upon doing
the differentiations in (\ref{eq:m2Qmu}) one gets\begin{equation}
m_{Q}^{2}(X)=2c\left(\frac{\alpha_{X}}{4\pi}\right)^{2}b'm_{3/2}^{2},\label{eq:m2Xamsb}\end{equation}
 which indeed would be negative if $b'$ is negative as is the case
for the $SU(2)\times U(1)$ group of the standard model. In other
words we would have the same problem as for AMSB!

However it does not really make sense to first take the limit and
then differentiate. $\mu$ is a mass scale and may be identified with
the vev of $X$ but not with the superfield itself. On the other hand
the formula (\ref{eq:m2Qmu}) makes sense only when $X$ is actually
treated as the full superfield before the differentiations, and then
set to its vev afterwards. Also since the scale $X$ is to be associated
with the Higgs vev and the mass scale of all other states (except
the top) are below this scale the limit should be taken from below
and before the $X$ differentiation as was done above to get (\ref{eq:m2X}).
So in approaching the SUSY breaking threshold from below it is probably
appropriate to take the running as being due to the standard model
states, though obviously the fact that some SUSY partners may well
be below the top quark makes the precise determination of this running
somewhat unclear. A detailed investigation of this will be left to
a future publication.

The same modification should be made for the DS calculation of gaugino
masses as well. Thus in (\ref{eq:HDS})(\ref{eq:maDS})(\ref{eq:maDS1})
the coefficient $b$ should be replaced by $b-b'$. It should also
be noted that although (\ref{eq:m2X}) is independent of the Higgs
vev the derivation requires the existence of a non-trivial minimum
for the Higgs potential. Since in the MSSM the symmetric vacuum is
only destabilized by radiative effects that depend on the breaking
of supersymmetry, this mechanism depends on a bootstrap like self-consistency
argument.

One also sees that (\ref{eq:m2X}) exhibits characteristics of both
AMSB and GMSB. Like the former the squared masses are proportional
to the squared gravitino mass. Furthermore like AMSB the classical
contributions to the mass splittings need to vanish since otherwise
they would dominate the quantum effects. This means that the supersymmetry
breaking sector needs to be sequestered \citep{Randall:1998uk}. An
example of how this could happen is the second model discussed in
section\ref{sec:Models-with-F^{C}=00003D00003D0}. By contrast in
GMSB the mass scale is set by the messenger mass, and the gravitino
is the lightest super-partner and one does not really need sequestering.
However unlike AMSB but as in GMSB this mechanism gives positive values
for all squared masses.

\section{Conclusions}

Let us summarize the main points of this paper.

\begin{itemize}
\item The expression for the anomaly mediated contribution to the gaugino
mass is essentially contained in \citep{Kaplunovsky:1994fg} and is
given here in equation (\ref{eq:gauginomass1}). When one uses the
value of the F-term of the Weyl compensator that is required to get
to the Einstein-Kaehler gauge, we get the formula (\ref{eq:gauginomass2})
which we claim is the correct formula for the gaugino mass that can
come purely from Weyl anomalies. This latter formula can alternatively
be derived without going to the Weyl compensator formalism (i.e. the
$C=1$ gauge) and in that case it comes from Jacobians associated
with field redefinitions that are associated with going to the Kaehler-Einstein
frame. 
\item An additional contribution to the gaugino mass comes from an effect
noticed in\citep{Dine:2007me} (DS). This when added to the previous
contribution gives a formula that is superficially similar to the
complete expression for the gaugino mass given in \citep{Bagger:1999rd}.
However the DS contribution can have additional terms, when there
is a {}``mu'' term in the MSSM superpotential for instance. 
\item There is no AMSB contribution to the soft masses. The usual argument
proceeds from inserting a conformal compensator superfield factor
$C$ to multiply the ratio of scales $\mu/\Lambda$ in the wave function
renormalization $Z(\mu/\Lambda)$. However this ratio, being a ratio
of physical scales should be independent of the Weyl gauge. Indeed
inserting such a factor will violate the Weyl invariance of the formalism
(which incidentally should be preserved whether or not there are mass
terms in the action). Furthermore any non-trivial dependence on $C$
in $Z$ will mean that the former becomes a propagating field which
cannot be decoupled from the action and would violate unitarity. In
any case one should be able to derive a physical effect in any gauge
- in particular in the usual formulation of supergravity with the
Weyl compensator superfield $C$ set to unity. This does not seem
to be possible - which again suggests that the effect, at least in
its original form, is absent. The point is that unless local supersymmetry
is explicitly broken by the regularization, one should be able to
express the Wilsonian effective action in terms of a superpotential
and an effective (quantum corrected) Kaehler potential in the standard
formulation of supergravity. 
\item There is however a contribution which is similar to the usual AMSB
one, that has been discovered by Dine and Seiberg \citep{Dine:2007me}.
However there are several differences. Firstly there is an additional
term when there is a {}``mu'' term present. Secondly we have shown
that there are models in which the usual AMSB contribution is zero
but the DS contribution is non-zero. Thirdly this DS contribution
does not give rise to negative slepton squared masses. Fourthly the
DS effect has nothing to do with Weyl anomalies and certainly exists
independently of the particular formulation of supergravity. 
\end{itemize}

\section{Acknowledgments}

I wish to thank Ramy Brustein, Kiwoon Choi, Michael Dine, Shamit Kachru,
Hans-Peter Nilles, Nati Seiberg, Fabio Zwirner and especially Vadim
Kaplunovsky for discussions and comments on the manuscript. This research
is supported in part by the United States Department of Energy under
grant DE-FG02-91-ER-40672.

\section*{Appendix}

It is helpful to consider some of the issues involved in the compensator
formalism and its relation to AMSB in a simplified context. Consider
the action (\ref{eq:action}) without gauge fields and with just one
matter field (say $Q$ with Kahler potential $K=\bar{Q}Q$) and the
Weyl compensator field $C$. Let us simplify further by taking these
fields to be real. The bosonic part takes the form of two conformally
coupled scalars and a potential term:\begin{equation}
S=\frac{1}{2}\int d^{4}x\sqrt{g}(C{}^{2}R+6g^{\mu\nu}\partial_{\mu}C\partial_{\nu}C)-\int d^{4}x\sqrt{g}[\frac{R}{6}C^{2}Q^{2}+g^{\mu\nu}\partial_{\mu}(CQ)\partial_{\nu}(CQ)+C^{4}V(Q)]\label{eq:S1}\end{equation}
This action has the Weyl invariance (descending from the super-Weyl
invariance of (\ref{eq:action})) \begin{equation}
g_{\mu\nu}\rightarrow e^{4\tau}g_{\mu\nu},\, C\rightarrow e^{-2\tau}C.\label{eq:W1}\end{equation}
Note that the Weyl compensator $C$ has a kinetic term with the wrong
sign. However this is not a problem since it can be guaged away -
it is really a spurious field which is equivalent to a Weyl transformation.
At the quantum level these transformations will have an anomaly with
the structure $\int\tau``R^{2}"$ where the integrand is a linear
combination of four derivative terms of the metric. This will need
to be cancelled by a similar term with $\ln C$ instead of $\tau$
that is added to the action so that the Weyl invariance is preserved
at the quantum level. This is of course essentially what we did in
section (\ref{sec:Weyl-Anomalies}) except that there we ignored squared
curvature terms and just focused on (supersymmetrized) gauge kinetic
terms. This is needed for consistency since we need to be able to
remove the spurious field $C$. The theory is completely equivalent
to that with the action (\ref{eq:S1}) in the gauge $C=1$.

The theory is however not in Einstein frame since the scalar fields
couple to curvature in the form $\int\sqrt{g}C^{2}(1-\frac{Q^{2}}{3})R$.
To go to the Einstein frame we simply pick the gauge $C=1/\sqrt{1-\frac{Q^{2}}{3}}$
and the action (ignoring the anomaly term) becomes \begin{equation}
S=\int d^{4}x\sqrt{g}[\frac{R}{2}-\frac{1}{(1-\frac{Q^{2}}{3})^{2}}(g^{\mu\nu}\partial_{\mu}Q\partial_{\nu}Q+V(Q))]\label{eq:S2}\end{equation}
Alternatively we could have started with the action (\ref{eq:action})
in $C=1$ gauge and then do a field redefinition (or equivalently
a Weyl transformation) $g_{\mu\nu}\rightarrow(1-\frac{Q^{2}}{3})^{-1}g_{\mu\nu}$.
It is easily checked that this leads to the same action as (\ref{eq:S2})
as it should. In the quantum theory this is not the whole story since
the field redefinition results in a Jacobian factor in the path integral
measure that results effectively in the same term as the one discussed
earlier. The main point is that the final action including the anomaly
correction must in fact be the same. $C$ is a spurious field and
can have no physical significance. This must be the case even if one
integrates the fluctuations of the field down from some scale $\Lambda$
(at which we take the the above action to be a valid description)
down to some lower scale $\mu$ at which we want to investigate its
physics. The corresponding renormalization constants can only depend
on the ratio of scales $\mu/\Lambda$ and clearly should not depend
on the spurious field $C$. Any such dependence would violate the
Weyl invariance which enabled us to decouple this field.

What is done in the literature on AMSB however is to break the Weyl
symmetry by picking a metric - i.e. the flat metric $\eta_{\mu\nu}$.
Once this is done of course one loses sight of the original invariance.
In flat space then the Weyl invariance is replaced by conformal invariance
which  is broken by mass terms. Thus let us define (as is usually
done in the literature) $\hat{Q}\equiv CQ$ and take the potential
to be $V=\lambda\phi^{4}$. The action (\ref{eq:S1}) then becomes
\[
S=\int d^{4}x\sqrt{g}[(C{}^{2}-\frac{\hat{Q}^{2}}{3})\frac{1}{2}R+3g^{\mu\nu}\partial_{\mu}C\partial_{\nu}C)-g^{\mu\nu}\partial_{\mu}\hat{Q}\partial_{\nu}\hat{Q}-\lambda\hat{Q}^{4}]\]
If one goes to flat space with the above metric it appears as if we
have a conformally invariant flat space theory for $\hat{Q}$ that
is independent of $C$. Any $C$ dependence would arise only if one
had explicit mass terms. However this ignores the fact that graviton
fluctuations will couple in a field dependent fashion and furthermore
that $C$ appears as a ghost. One needs to go to the Einstein frame
by doing a field redefinition $g_{_{\mu\nu}}\rightarrow(C^{2}-\hat{Q}^{2}/3)^{-1}g_{\mu\nu}$
and it is the new metric that should be put equal to the flat metric.
This transformation however introduces $C$dependence into the $\hat{Q}$
lagrangian - the potential for example becomes $\lambda\hat{Q}^{4}/(C^{2}-\hat{Q}^{2}/3$). 

In any case the issue is not conformal invariance. What is relevant
is Weyl invariance which exsits irrespective of the existence of mass
terms. It is this invariance (which is manifest only if the metric
is not fixed) that enables one to eliminate the spurious field $C$.

\bibliographystyle{apsrev}
\bibliography{myrefs}

\end{document}